# Forecasting and skipping to Reduce Transmission Energy in WSN

**Ahmad A. ABBOUD[1], Abdel-Karim Yazbek[2], Jean-Pierre Cances[3], Vahid Meghdadi[4]**

[1,2,3,4]Department of Components Circuits Signals and High Frequency Systems
Xlim Labs, University of Limoges
Limoges, France
Ahmad.Abboud @etu.unilim.fr, {Abdel-karim.yazbeck, Cances,Meghdadi} @ensil.unilim.fr

**Abstract**

This paper deals with the improvement of energy efficiency in wireless sensor networks (WSN).Taking into consideration the power saving problem which is of crucial importance when the sensors are supplied by a limited power source, this paper proposes a method that optimizes as much as possible the transmission power of the sensors. Under the assumption of perfect channel between the Base Station (BS) and the Sensor Nodes (SN's) and with sufficient power resources at the BS, transferring the effort of transmission power to the BS will not be a peculiar issue. This paper proposes a method that reduces the transmitted data at the SN's and compensate this by requesting the variance of the measured value form predicted values. Furthermore, a request management algorithm (RMA) is developed to reduce the amount of requested data based on consecutive successive predictions.

The result of this method reveals a major impact on reducing the transmission power at the SN's.

## 1 INTRODUCTION

Demands on data measurement from various fields including environmental parameters (Temperature, Pressure, Humidity, etc …) and other health care and military issues, lead to consolidate the research on WSN's.

Among many challenges facing this technology, power saving attracts many researchers to work on. Of course this is due to the limited power resource that serves the SN's under hard environmental circumstances that limits the possibility to recharge. As the processing unit of the SN's is classical the most power consumption unit of the system resides in the transmission process. And, as the size of the data increases, the power consumption will follow it linearly.

Many researches on saving the power of SN's by reducing the size of the transmitted data were done and a convenient results had been presented, but this proposal can prove its efficiency in special cases where measured values are relatively close to each other, and/or follow a trivial cycle as in (daylight, temperature, humidity along a day, etc…).

The key concept behind the proposed technique relies on the fact that BS can learn from previous patterns how to predict an approximate value about the new measurement before requesting it from the sensors. This will limit the content of the replied packet (from the sensors)by just including the error correction, or the variance from the predicted value received in the request packet. In case of the predicted value is approximately equal to the measured value, SN may not reply and this will be held as a zero error. BS only stores the predicted value. As an initial case BS must send a predicted value equal to zero which will oblige the sensor to send the measured value.

As long as the error value is small compared to the measured one this will lead to less bits to be transmitted and this depends directly on the accuracy of the prediction.

### 1.1 Related Works

A proposed method based on multivariate spatial and temporal correlation to improve prediction accuracy in data reduction was presented in [1].The proposal outperforms some current solutions by about 50% in humidity prediction and 21% in light prediction.

Another research was done on prediction based technique [2] and showed that it was possible to conserve a significant amount of energy through the proper use of data prediction and node scheduling without a significant loss in accuracy. Their results show that it is possible to increase lifetime by up to 2600% at the cost of increasing average error by $0.5C^0$ for temperature or 1.5% for humidity measurements.

Authors in [3] have developed data reduction strategies which reduce the amount of sent data by predicting the measured values both at the source and the sink, requiring transmission only if a certain acceptance error differs by a given margin from the predicted values. Their technique is based on using variable step size LMS algorithm. They had achieved maximum data reduction of over 95%, while retaining a reasonably high precision.

## 1.2 Paper Organization

The rest of the paper will be managed as following. At first, a clear definition of the system structure and postulates will be presented. The responsibility and the operation algorithms of the BS and SN will be presented in section 3 and 4 respectively. Then in section 5,the ANN forecasting method used to predict future patterns will be discussed. In section 6 packet reciprocity between BS and SN's will be presented and then, in section 7, a novel algorithm used to manage requests at the BS will be exploited to reduce the request frequency. At last numerical results and conclusion will be presented in Section 8 and 9 respectively.

## 2   SYSTEM STRUCTURE

The system in which to apply the proposed technique is assumed to include a base station connected to a large number of wireless sensors. Correlation between sensors is applicable in this model, unless for simplicity, we deal with it as an uncorrelated system, where the communication is hold directly with the base station. Figure 1 represents a simple network where all sensors are connected directly to the BS, like for example in the SIGFOX system.

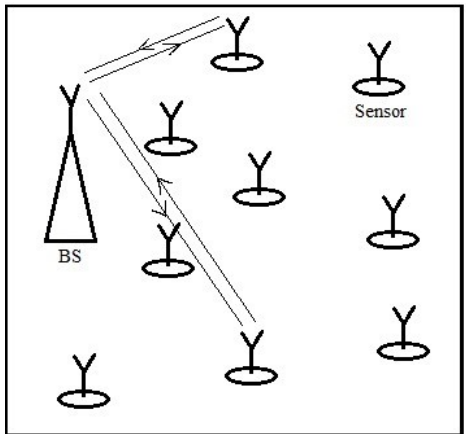

Figure 1: WSN structure

In order to focus on the subject perfect channel links between BS and SN's were assumed.

The listed model is simple, basic, scalable and interoperable where data aggregation, multi-layer structure and many other models are able to extend the system without losing any of its perfections.

## 3   BSOPERATION &RESPONSIBILITY

In order to be able to collect data from all SN's and then estimating the measured values, BS must have a sufficient energy source and a classical processing unit with a large storage device. BS is responsible to request data from sensors in a consecutive way. The key behind requesting the data from the sensors, and not to wait the sensors to send their data freely, is to reduce the transmission process at the sensors, especially when the base station has a good prediction about the measured data. The requested packet contains a predicted value **E**, an accepted error value **α** and other fields related to datalink and sequence number. The predicted value **E** is calculated based on previous knowledge learned from stored patterns, and **α** is the accepted error in which measured value **M** is supposed to be approximately equal to **E** if the following inequality holds.

$$M - α ≤ E ≤ M + α \quad (1)$$

**α** can be selected according to the demanding data accuracy, where small **α** values lead to higher accuracy.

After requesting a measured value from a SN, BS will wait a specific duration **T** until the sensor responds. In case of no response from the sensor, the predicted value **E** is supposed to be approximately equal to the measured value and it will be stored in the BS as if it was a collected data. In case of response the received packet must contain a signed variance value from the prediction**V**, which is supposed to fit the following inequality

$$\log_2(V) ≤ \log_2(E) \quad (2)$$

The stored value **S** at the BS will be expressed as

$$S = E + V \quad (3)$$

$S_{tk}$ represents the stored measured value from the k*th* SN at time t*th* period. Note that SN's are separated in columns and rows for geographical distribution issue. The operating algorithm at the BS is listed below:

**RequestMeasure( E , α)**

1. **Request(E=0,α);**
2. **Wait(T);**
3. **V=Received( );**
4. **S$_t$=E$_t$+V;**
5. **E$_t$=Predict ( S$_{t-n}$, S$_{t-1}$ );**
6. **Request ( E$_t$ ,α );**

7. Goto 2

It's clear from the above algorithm that the initial predicted value is zero where BS hasn't got any previous knowledge to predict the future values. At step 3, while waiting a period **T** if there is a received packet, then **V** can be extracted and added to the predicted value. Else **V** will be equal to its initial value and **S=E**, then a prediction based on previous stored data will take place for the next iteration.

## 4 SENSOR OPERATION &RESPONSIBILITY

According to the proposed model, sensor nodes are simple devices that must contain just a sensing tool with analog subtractor, comparator and a transceiver unit supplied by a small power source (see figure 2). As mentioned before this model is scalable, thus extensions are easily implemented in order to obtain constructive results.

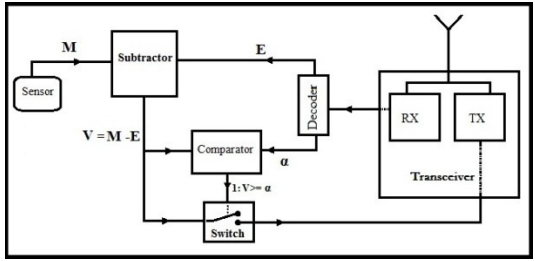

Figure 2: Sensor Node Block Diagram

The operation of the sensor is simplified in order to reduce computational and transmission power consumption as much as possible. After receiving a request packet from the BS, sensors tend to subtract the received Predicted value from the measured value.

$$V = M - E \quad (4)$$

Then **V** is compared to the accepted error value **α**. If |**V**| > α the sensor will transmit **V**, else no transmission will take place.

The following algorithm presents the operation.

1. ReceiveRequest(E , α );
2. M= MeasureValue()
3. V = M – E;
4. If (V < - α OR V > α)
5. Send(V);
6. Goto 1

From the previous algorithm it's clear that the sensor will never send any value unless **V** is not within the acceptance error range.

## 5 FORECASTING USING ANN

As to minimize the error between measured sensor data and predicted data, we have to choose the best performance forecasting technique to solve the problem. Taking into advantage the slow change in environmental measurements where in most cases they are changing slowly and periodically (daily solar radiation, temperature, seasonal rain falling, pressure, humidity, etc …), forecasting algorithm works well and especially time series forecasting.

Many algorithms can be applied to forecast time series data, includes ARIMA (Auto Regression Integrated Moving Average) [4], Bayesian interference [5],[6], Markov chains [7],[8], K-Nearest Neighbors algorithm [9],[10] and ANN (Artificial Neural Networks) [11], [12]. In this paper we apply ANN to predict future measurements and this choice is based on many research works that shows the ability of ANN to predict time series data [13], [14],[15]. Moreover, ANN provides a non-linear parametric model compared to classical methods.

ANN is an artificial intelligence system that is composed of simple processing units called neurons, each neuron is connected to its neighbor with a weighted links that are able to be updated during learning epochs. The stored knowledge acquired by the network can be characterized by the state of the weights of all the links. Neurons are composed of input connections collected to a node that has an activation function and a threshold value, where the last decides if the traversed data must pass to the output connection or not.(see figure 3 below)

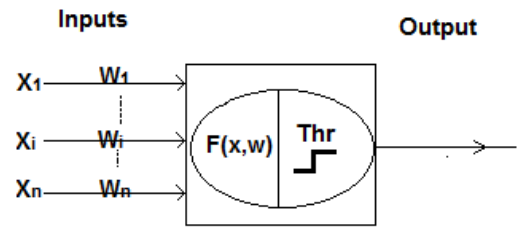

Figure 3:Artificial Neuron

In our work dynamic neural networks, which include time delay lines, are used to perform non-linear autoregressive prediction with external time step input vector. This type of networks contains multi-hidden layers of neurons with delay factor connected to an output layer. The training of such network can be done using several algorithms e.g. (Bayesian Regularization,

3 | P a g e

Levenberg-Marquardt, Scaled Conjugate Gradient.), where Bayesian Regularization algorithm is proffered to be used with noisy input data, unless it will take more processing epochs.

The network must be trained with past targets that were previously collected by the BS. After training the ANN, a closed loop will be performed by connecting the output of the ANN to its target input. The new output will be considered as a series of prediction values that will be sent by the BS to the SN within the request message.

As a result the prediction values within acceptable error rate when compared to measured values will save the transmission energy of the SN. This leads to say that the performance of the forecasting method will directly affect the performance in power saving at the SN.

## 6  PACKET RECIPROCITY

To maintain date exchange in any communication system, a control channel which collaborates with the data channel is necessary. Because of the simplicity of the SN the major tasks must be hold by the BS. As a control message protocol, BS can test the connection state between any of its SN's by simply sending two consecutive requests. The first request contains the following parameters, (**E**=0 and **α**=0). Evaluating in (4) results in an expected outcome which equals exactly the measured value M.

If **M** is also equal to zero, then the second request packet can assure this by assigning a large unexpected value to E that can't be reached by the sensor rapidly and keeping **α**=0. If the sensor reply with approximately the negative value of **E** then **M**=0.

In order to control the flow of data BS can decrease the frequency of the requested packet as much as the sensor never responds (see figure4).

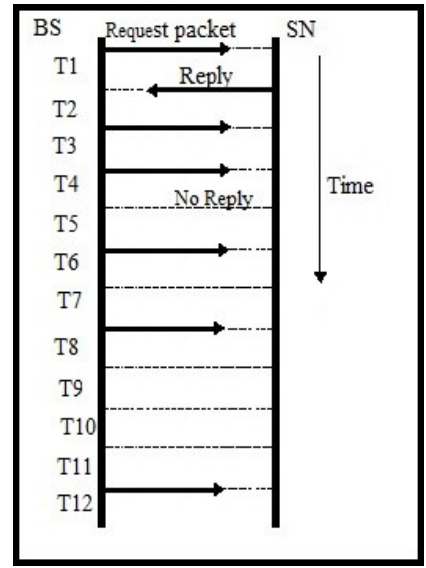

Figure 4: Packet Exchange

## 7  REQUEST MANAGEMENT ALGORITHM

Under the assumption of optimal prediction, a lot of requested packets will never been replied. This not due to an imperfection in channel link but this the consequence of the perfect prediction. Even if a SN never replies it has already consumed some power at the receiving phase and at the processing phase. In order to exploit the power source as much as possible, requested packet frequency must be controlled in a manner that insures data accuracy.

A method acquired from TCP congestion control can be recruited to control the flow of requested packets in WSN. Let **C**∈ℕ, defined as the number of consecutive non replied packet which indicate a perfect prediction. Let **q** ∈ ℕ, defined as the number of iteration to be skipped before the next requested packet.

We define three phases of packet request growth:

1- Classical Phase (CP): where requested packet are retrieved consecutively for each $S_t$ which indicates **q**=0.
2- Fast State Phase (FSP): where requested packets growth exponentially and thus the next iteration will increase double the last increment, which indicates
3- **q**= **q**x2.

4- Linear State Phase (LSP): where the requested packet growth linearly by incrementing the growth of iteration number by 1, which indicates **q=q +1**.

Request Management Algorithm (RMA) at the base station will operate as follows:

```
1.  Start
2.  Eₜ=Predict ( Sₜ₋ₙ, Sₜ₋₁ )
3.  Request ( Eₜ ,α )
4.  while ( T )
5.          Sₜ = Received_V ( ) + Eₜ

6.  if ( Received_V ( ) = 0 )
7.          Sₜ = Received_V ( ) + Eₜ
8.          Promote ( C )
9.  else    Degrade ( C )

10. if ( C ≥ Tr2 )
11.         Fill_Skips ( t till t +2 )
12.         t = t + 2
13. else if ( C ≥ Tr1 )
14.         Fill_Skips ( t till t ×2 )
15.         t = t × 2

16. else    t = t + 1

17. Goto    Start
```

Where the **Fill_Skips( )** function represented below:

```
1.  Fill_Skips ( x till y )
2.      for ( k = x till y )
3.              Sₖ = Predict ( Sₖ₋ₙ, Sₖ₋₁ )
```

This algorithm is similar to the TCP congestion control algorithm in some manners, but the most significant issue is that there are two thresholds **Tr1** and **Tr2** which are defined as the number of consecutive non replied packet that separates FSP and LSP respectively.

When assigning a large number that can't be expected to the first threshold **Tr1**, the performance of this algorithm will be reduced to the previous classical one in mentioned in section 3. In this case no skipped Packet Requests will take place and the system will work on classical phase.

In case of a series of consecutive perfect predictions **C ≥ Tr1** the numbers of skipped requested packets **q** will grow exponentially and this is called the FSP. After that phase, if **C** continues to increase till it reaches **Tr2** then **q** will grow linearly to avoid data loss.

Figure 5 presents the shape of the 3 phases and the growth of skips with respect to the iteration number.

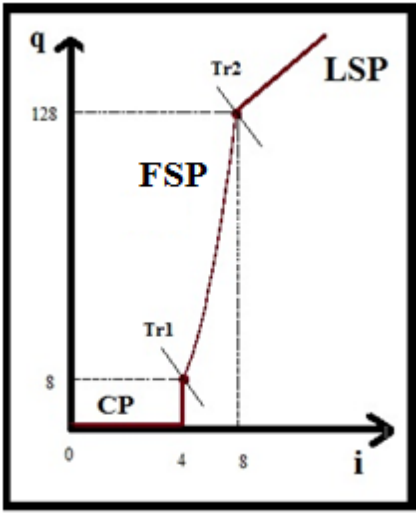

Figure 5: number of skips ( q ) relative to the iteration number ( i ).

In case of any error variance replied from a sensor then one backward phase will take place.

Note that, skipped requests are compensated by predicted values, which will save the power of receiving and processing data at the SN.

## 8 NUMERICAL RESULTS

In order to evaluate the performance of the system, we introduce the reciprocity of one sensor with the Base station where the BS must request measured data every **T**=1 hr, the requested data are the measurements of temperature for 400 hr. The forecasting algorithm was trained for 250 hr and forecasting start takes place for the rest of the experiment. Figure 6 shows 150 hr of predicted temperature and expected measurements.

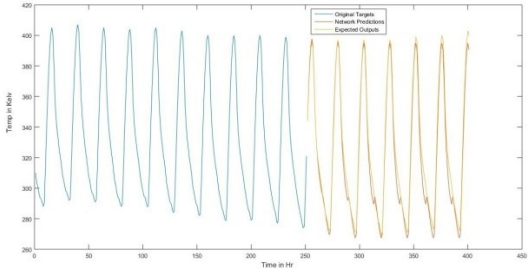

Figure 6 predicted patterns Vs Expected

This result is obtained by an ANN of 50 hidden layer and one output layer trained using Bayesian Regularization Algorithm with 250 target input. The structure of this network is shown in figure 7.

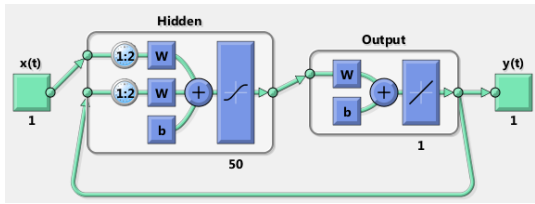

Figure 7 ANN structure

Where x(t) presents the input target vector and y(t) presents the output vector. After 1000 epoch on training the network by 250 input target a feedback connection was connected from output to input thus prediction of next patterns is possible. The network is asked to predict 150 patterns, where the performance of the network is shown in figure 8.

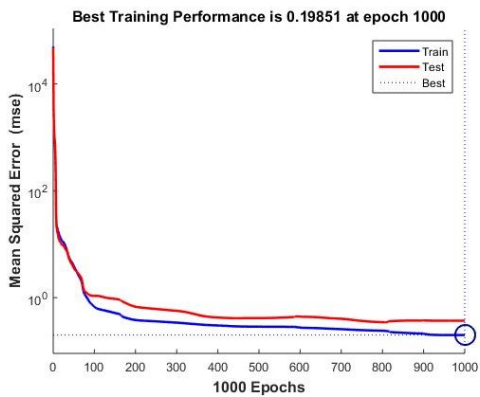

Figure 8 Performance

The error histogram after training the network is presented in figure 9.

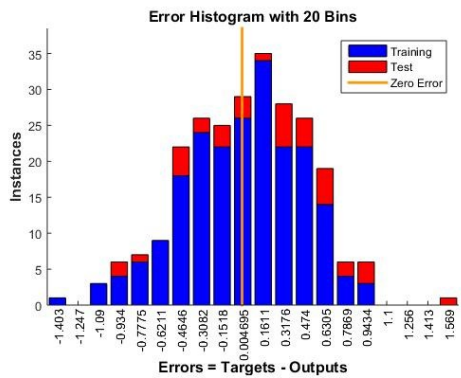

Figure 9 Error Histogram

The regression after training the network is presented in figure10.

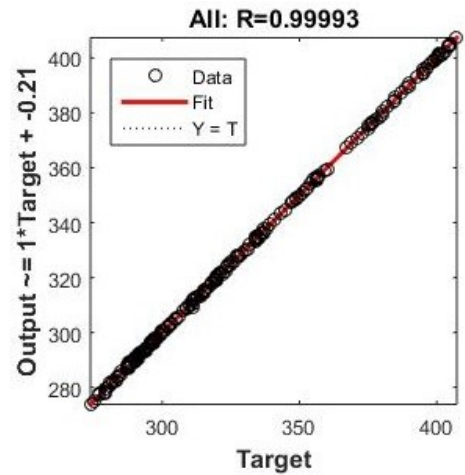

Figure 10 Data Regression

It is clear from the regression plot that the output data well fit the target with a regression of 0.99993.

After applying the request management algorithm on the BS a plot shows the stored data and the actual measurements is presented in figure 11 where $\alpha$=1, Threshold1=3 and Threshold 2= 7.

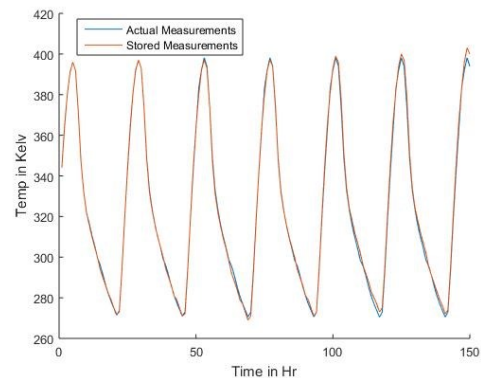

Figure 11 Stored Vs Actual measurement

The stored values can replace the actual measured values within an error of 1degree. Among the 150 predicted patterns the number of skips is 132 and the total number of accepted prediction to be stored is 139 patterns.

This results leads to a reduction in the computational energy cost at the sensor by approximately 833% and a reduction of transmission energy of approximately 1363% at the sensor.

## 9   CONCLUSION

The results obtained are promising and the forecasting techniques with request management algorithm prove their ability to reduce energy consumption at the SN with credible amount. We can conclude from the results that the performance of the forecasting algorithm and the type of data being measured are the main factors of reduce the energy consumption of the SN. Another advantage of this algorithm is its ability to be implemented additively with most WSN architectures without contradicting the performance of the system.

Future research will be done on the ability of RMA to be applied on MWSN where plenty of pattern can be predicted and high precision are not a priority.

**Biographical notes:**

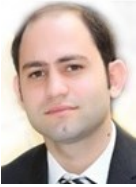 **¹Ahmed A. Abboud** (2$^{nd}$ Oct 1985) was born in MazratMichref South Lebanon. Received a technical diploma in communication & computer engineering from University Institute of Technology (UIT) Jwaya, South Lebanon and the Master of Science in Computer Science & Communication from Arts, Sciences & Technology University in Lebanon, in 2008 and 2012 respectively. He is now a PHD student at the University of Limoges since October 2014. His research interests are in the area of applying artificial intelligence algorithms on MIMO communication systems.